\documentclass{elsart}
\begin{document}
\begin{frontmatter}
\title{ Vortex Matter in Mesoscopic Superconducting Disks and Rings}

\author
{F.M. Peeters \cite{o:gnu}, V.A. Schweigert \cite {*:gnu}, B.J. Baelus and 
P.S. Deo \cite {f:gnu}}
\address{\it  Departement Natuurkunde, Universiteit Antwerpen (UIA),\\
Universiteitsplein 1, B-2610 Antwerpen, Belgium}
\date{\today}
\maketitle

\begin{abstract}
Phase transitions between different (i.e. giant and multi-vortex)
superconducting states and between the
superconducting-normal state of mesoscopic disks and rings are studied 
in the presence of an external magnetic field by
solving the two non-linear Ginzburg-Landau equations self-consistently. 
The flux through a circular disk with a hole in the middle
is not quantized. 
 
\end{abstract}

\end{frontmatter}

\section{Introduction}
 
Modern fabrication technology has made it possible to 
construct superconducting samples of micro- and submicrometer 
dimensions\cite{mos1,geim} which are comparable to the important length scales
in the superconductor: the coherence ($\xi$)
and the penetration length ($\lambda$). In such samples classical finite size 
effects are important which leads to flux confinement.
The dimensions of the sample are considered to be sufficiently large such that
quantum confinement of the single electron states are neglectable and 
the superconducting gap is not altered by the confinement. 
This is the regime in which the Ginzburg-Landau (GL) theory is expected
to describe the superconducting state.

The size and shape of such samples 
strongly influences the superconducting properties. 
Whereas bulk superconductivity exists at 
low magnetic field (either the Meissner state for $H<H_{c1}$ in type-I and 
type-II
superconductors or the Abrikosov vortex state for $H_{c1}<H<H_{c2}$
in type-II superconductors), surface superconductivity survives
in infinitely large bounded samples up to the third critical
field $H_{c3}\approx 1.695H_{c2}$\cite{saint}.
For mesoscopic samples of different shape,
whose characteristic sizes are comparable to the coherence
length $\xi$, recent experiments \cite{mos1,geim,buisson,strunk1,zhang}
have demonstrated the Little-Parks-type \cite{little}
oscillatory behavior of the phase boundary between the normal
and superconducting state in the $H-T$ plane, where $T$ and $H$ are the
critical temperature and the applied magnetic field, respectively.

In the present paper we study superconducting disks of finite height with a
hole in the center. 
Previous calculations were mainly concentrated on circular infinite long
superconducting wires and hollow cylinders.
In this case there are no demagnetization effects. 
Superconducting disks and rings of finite thickness have been studied 
much less extensively theoretically although they may often correspond more closely to 
experimental fabricated systems. 

The giant vortex state in cylinders was studied in Ref. \cite{mosh97}
and in Ref. \cite{benoist} for infinitely thin disks and rings. 
Recently, Venegas and Sardella \cite{venegas} used the London theory to study
the vortex configurations in a cylinder for up to 18 vortices. They found that
those vortices form a ring structure very similar to classical confined
charged particles \cite{bedanov}. These results are analogous to those of Refs.
\cite{buzdin,brongersma} where the image method was used to determine the
vortex configuration. 
The transition from the giant vortex to the multi-vortex state in a thin
superconducting disk was studied in Ref. \cite{schweigert98}.
Square-shaped loops were investigated by Fomin {\it et al} \cite{fomin} in order to
explain the Little-Park oscillations found by Moshchalkov {\it et al}
\cite{mos1} and to study the effect of leads. In this case the
magnetic field was taken uniform along the z-direction and demagnetization
effects were disregarded. 
Loops of non zero width were studied by Bruyndoncx {\it et al} \cite{bruyndoncx}
within the linearized GL theory. The magnetic field was taken uniform and equal
to the applied field which is valid for very thin rings. They studied the
$T_c(H)$ phase boundary and in particular the 2D-3D dimensional cross over
regime.
A superconducting film with a circular hole was studied by
Bezryadin {\it et al} \cite{bezryadin}. This is the limiting case of the
current ring structure for wide rings. Or conversely, it is the antidot version
of the disk system studied in Refs. \cite{geim,deo}.

Most of previous calculations are restricted to the London limit (i.e. extreme
type-II systems) or do not consider the demagnetization effect. 
Our approach is
not limited to such situations and we calculate the superconducting
wavefunction and the magnetic field fully self-consistently for a finite 3D
system, i.e. for a disk and ring with finite width and thickness. 
The details of our numerical approach can be found in Ref. \cite{me}
and therefore will not be repeated. 
Here we investigate the flux expulsion in the Meissner state,
the giant $\rightarrow$ multi-vortex
transition in fat rings, 
and the absence of flux quantization in rings of finite height and
non zero width. 
In contrast to most of earlier work (see e.g. \cite{bruyndoncx}) we are
interested in the superconducting state deep inside the $T_c(H)$ phase diagram
where the magnetic field in and around the superconductor is no longer
homogeneous. The transition from disk to thin loop structures is also 
investigated.  

\section{Meissner state}

Meissner and Ochsenfeld (1933) found that if a superconductor is cooled in a
magnetic field to below the transition temperature, then at the transition the
lines of induction are pushed out. The Meissner effect shows that a bulk
superconductor behaves as if inside the specimen the magnetic field is zero,
and consequently it is an ideal diamagnet. The magnetization is
$M=-H_{applied}/4\pi$. This result is shown by the dashed curve in Fig. 1
which refers to the right axis.

Recently, Geim {\it et al} \cite{geim} investigated superconducting disks of
different sizes. Fig. 1 shows the experimental results (symbols) for an Al disk of
radius $R \sim 0.5 \mu m$ and thickness $d \sim 0.15 \mu m$. 
Our numerical results which includes a full self-consistent solution of the two
non-linear GL equations are given by the solid curve but which refers to the
right axis (we took $\lambda(0) = 0.07 \mu m$, $\xi(0) = 0.25 \mu m$ and
$\kappa = 0.28$). There is very good qualitative agreement but quantitatively the
theoretical results differ by a factor of $50.5$. The latter can be understood
as due to a detector effect. Experimentally the magnetization is measured using
a Hall bar. Consequently, the magnetic field is averaged over the Hall cross
and it is the magnetization resulting from the 
field expelled from this Hall cross which is plotted while
theoretically we plotted the magnetization resulting from the 
field expelled from the disk.
This is illustrated in Fig. 2 where we show a contour plot of 
the magnetic field profile in the plane of the middle of the disk 
and a cut through it (see inset of Fig. 2). The results are for a disk of radius $R=0.8
\mu m$ for a magnetic field such that $L=5$ which implies that a giant vortex
with six flux quanta is centered in the middle of the disk leading to an
enhanced magnetic field near the center of the disk
and expulsion of the field in a ring-like
region near the edge of the sample. The demagnetization effect is clearly
visible which leads to a strong enhancement of the field near the outer
boundary of the disk. The theoretical result in Fig. 1 corresponds to an averaging of
the magnetic field over the disk region while the Hall detector averages the
field over a much larger area which brings the average field much closer to the
applied field. The inset of Fig. 2 shows clearly
regions with $H < H_{applied}$ which correspond to diamagnetic
response and regions with $H > H_{applied}$ which correspond to paramagnetic
response. The averaging over the detector area (if width of the detector
$W > R$) adds additional
averaging over paramagnetic regions to the detector output.  

Using a Hall bar of
width $W=2.5\mu m$ a distance $h=0.15 \mu m$ separated from the
superconducting disk results into an expelled field which is a factor $50.44$
smaller than the expelled field of the disk and brings our theoretical results
in Fig. 1 in quantitative agreement with the experimental results.
This averaging over the Hall cross
scales the results but does not change the shape of the curve as long as $L$ is
kept fixed. For different $L$ this scale factor is slightly different
because it leads to different magnetic field distributions. 

Notice also that the magnetization as function of the magnetic field is linear
only over a small magnetic field range, i.e. $H<20 G$
and the 
slope (see dashed curve in Fig. 1) is a factor 2.5 smaller than
expected from an ideal diamagnet. This clearly indicates that for
such small and thin disks there is a substantial penetration of the magnetic
field into the disk. This is illustrated in Fig. 3 where the magnetic field
lines are shown for a superconducting disk of radius $R=0.8\mu m$ in the $L=0$
state, i.e. the Meissner state. 
This strong penetration of the magnetic field inside the disk is also
responsible for the highly nonlinear magnetization curve for $H \gg 20 G$. 

\section{Giant-vortex state versus multi-vortex state}

Next, we generalize our system to a disk with thickness $d$ and 
radius $R_o$ containing a hole in
the center with radius $R_i$. 
The system under consideration is circular symmetric and therefore if the GL
equations would have been linear 
the Cooper pair wavefunction could have been written as
$\Psi(\vec{\rho})=F(\rho)\exp(iL\phi)$. We found that even for the nonlinear
problem such a circular symmetric state, also called giant vortex state,
still has the lowest energy when
the confinement effects are dominant.
This is the case when $R_o$ is small or $R_o/R_i \approx 1$, or in the case of large
magnetic fields when there is only surface superconductivity.
This is illustrated in the phase diagram shown in Fig. 4 for a thin
superconducting ring of outer radius $R_o/\xi=4$ and
width $R_o-R_i$. The equilibrium regions for the giant vortex states with
different angular momentum $L$ are separated by the solid curves. 
Notice that with decreasing width,
i.e. increasing inner radius $R_i$: 1) the superconducting state survives up to
large magnetic fields. This is a consequence of the enhanced superconductivity
due to surface superconductivity which in the limit of $R_i \rightarrow R_o$ 
leads to $H_{c3} \rightarrow \infty$ \cite{schweigert99}. 
2) The L-transitions occur for $\Phi = (L+1/2)\Phi_o$ 
in the limit $R_i \rightarrow R_o$, where $\Phi_o=hc/2e$ is the flux quantum.

For large type-II systems one expects the giant-vortex state to break up into
the Abrikosov triangular lattice of single vortices. The nonlinear term in the
GL theory is responsible for this symmetry breaking. 
Such a multi-vortex state is the lowest energy state for the shaded areas in Fig.
4. Notice that as compared to the disk case the presence of the hole in the
center stabilizes the multi-vortex state. Except for large $R_i$ because then
the confinement effect starts to dominate which imposes the symmetry of the
edge of the system on the superconducting state. Near the
normal/superconducting boundary only surface superconductivity survives and the
symmetry of the superconducting state is determined by the symmetry of the
surface which leads to the giant vortex state. The same holds for the Meissner
state (i.e. $L=0$). For $L=1$ there is one flux quantum in the center of the
system and there is no distinction between the giant and the multi-vortex
states. Only for $L \ge 2$ we can have broken symmetry configurations. 

The transition from the multi-vortex state to the giant-vortex state with
increasing magnetic field is illustrated in Fig. 5 where we plot the 
superconducting electron density for a disk with radius
$R_o/\xi=4$ containing a hole in the center with radius $R_i/\xi=0.4$
and for a magnetic field range such that the vorticity is $L=4$.
First, we can clearly discriminate three vortices arranged in a triangle with
one vortex in the center piercing through the hole (solid circle) of the disk. 
Increasing the magnetic field drives the vortices closer to the center and 
the single vortices start to overlap. Finally, for sufficiently 
large magnetic fields the four vortices form one circular symmetric giant-vortex state
with winding number $L=4$.

If the hole is sufficiently large more than one flux quantum can 
pierce through this hole for sufficiently large magnetic fields.
This is illustrated in Fig. 6 for a ring of $R_o/\xi=4, R_i/\xi=1$ with an external
magnetic field such that the
vorticity is $L=5$. There are four vortices in the superconducting material
arranged on the edge of a square and two piercing through the hole. 
The latter conclusion can be easily verified by considering 
a contour plot of the phase (right contour plot in Fig. 6). Notice that
encircling the hole the phase of the superconducting wavefunction
changes with $2\times 2\pi$ while encircling the outer edge of the ring it
changes with $6\times 2\pi$. In this plot the location of the vortices is also
clearly visible. Encircling a single vortex changes the phase by $2 \pi$.

\section{Is the flux through the hole of the ring quantized?}

It is often mentioned that the flux through a loop is quantized into
integer multiples of the flux quantum $\Phi_o=hc/2e$. 
In order to check the validity of this 
assertion we plot in Fig. 7(a) the
flux through the hole of a wide ring of thickness $d/\xi=0.1$
with outer $R_o/\xi=2$ and inner radius $R_i/\xi=1$. From this figure 
it is clear that the flux through the hole of such a ring is not quantized.
Note that depending on the magnetic field strength there is
compression or expulsion of magnetic field in the hole region 
(compare solid curve with dashed line which represents the flux $\Phi=\pi R_i^2H$).
The absence of quantization was also found earlier \cite{groff}
for hollow cylinders when
the thickness of the cylinder wall is smaller than the penetration length of
the magnetic field.
In order to understand this apparent breakdown of flux quantization lets 
turn to the derivation of the flux quantization condition. Inserting the
wavefunction $\Psi = |\Psi|\exp(i\phi)$ into the current operator we obtain
(under the assumption that the spatial variation of the superconducting density
is weak)
\begin{equation}
\vec{j}= \frac{e\hbar}{m}|\Psi|^2(\vec{\nabla} \phi - \frac{2e}{\hbar
c}\vec{A}) ,
\end{equation}
which after integrating over a closed contour C inside the superconductor
leads to
\begin{equation}
\oint_C \left( \frac{mc}{2e^2|\Psi|^2} \vec{j} - \vec{A} \right) \cdot 
d\vec{l} = L\Phi_o .
\end{equation}
When the contour C is chosen along a path such that the superconducting current
is zero we obtain
\begin{equation}
L \Phi_o = \oint_C \vec{A} \cdot d\vec{l} = \int rot{\vec A} \cdot d{\vec S} = \int
{\vec H} \cdot d{\vec S} = \Phi ,
\end{equation}
which tells us that the flux through the area encircled by C is quantized. 
In our wide superconducting ring the current is non zero at the inner boundary
of the ring and consequently the flux through the hole does not have to be
quantized. In fact the flux will be $\Phi = L\Phi_o + \Phi_j$ where
\begin{equation}
\Phi_j = \Phi_o \frac{m}{he} \oint_C \frac{\vec{j}}{|\Psi|^2} \cdot d\vec{l} ,
\end{equation}
which depends on the size of the current at the surface of the inner ring.  

The radius of the considered ring in Fig. 7 is sufficiently small that we have
a circular symmetric vortex state and consequently the current has only an azimuthal
component. The radial variation of this current is plotted in Fig. 8(a)
for three different values of the winding number $L$ for a fixed magnetic field. For $L
\ne 0$ the current in such a wide ring reaches zero at some radial position
$\rho^*$. As shown in Fig. 8(b) the corresponding flux through a surface with
radius $\rho^*$ is exaclty quantized into $L\Phi_o$
as required by the above quantization
condition. The current at the inner edge of the ring is clearly non zero and
therefore it is easy to understand why the flux through the hole is not
an integer multiple of the flux quantum. The effective radius $\rho^*$ is plotted in
Fig. 7(b) which is an oscillating function of the magnetic field. The dashed
curves in the figure correspond to the values for the metastable states. Notice
that $\rho^*$ oscillates around the value $\sqrt{R_oR_i} = \sqrt{2} = 1.41$
which can be obtained within the London theory \cite{zharkov}.  
At the $L \rightarrow L+1$ transition this radius exhibits a jump which moves
closer towards the outer radius while for fixed $L$ the effective radius
becomes closer to the inner radius with increasing $H$.

Next we consider the magnetic field range, $\Delta H$, 
needed to increase $L$ with one flux quantum, i.e. it
is the distance between two consecutive jumps in the magnetization. 
The corresponding flux $\Delta \phi = \pi R^2_o \Delta H$ is shown 
in Fig. 9. For small $R_o$ (see Fig. 9(a)) this value is almost independent
of $L$ for fixed $R_i$. Its value is approximately given by the flux
increase with one flux quantum through a circular area with radius $\sqrt{R_oR_i}$
(dashed horizontal curves in Fig. 9). 
For larger $R_o$ (see Fig. 9(b)) this relation is much more
complicated and the distance
between the jumps in the magnetic field is a strong function of
$L$, except for $R_i \sim R_o$ where it is again determined by the flux
quantization through an area with radius $\rho^* = \sqrt{R_iR_o}$. 

\section{Conclusion}

Mesoscopic superconducting disks of non zero thickness containing a hole in the
center were studied theoretically by solving numerically the coupled non linear
GL equations. In thin structures there is a substantial penetration of the
magnetic field into the superconductor leading to a large (non linear) demagnetization
effect. The hole in the disk enhances superconductivity and for small holes it
stabilizes the multi-vortex state. The flux through the hole of the disk is not
quantized. But for rings with a narrow width an increase of the applied
magnetic field with one flux quantum through a region with radius
$\sqrt{R_iR_o}$ increases the winding number $L$ with one unit.

{\bf Acknowledgement}
We acknowledge discussions with A. Geim and V. Moshchalkov.
This work was supported by the
Flemish Science Foundation (FWO-Vl) and IUAP-VI. FMP is a research director
with the FWO-Vl.

\newpage

\begin{figure}
\caption{Magnetization of an Al disk with radius $R=0.44 \mu m$ as function of
the applied magnetic field. The symbols are the experimental results and the
solid curve is our theoretical result.
}
\label{fig1}
\end{figure}

\begin{figure}
\caption{
The magnetic field distribution in the horizontal plane through the middle of
the superconducting disk with radius $R=0.8 \mu m$ for $L=6$. The inset gives
the magnetic field profile along the radial direction. 
}
\label{fig2}
\end{figure}

\begin{figure}
\caption{
The magnetic field distribution in and around a superconducting disk with
radius $R=0.8 \mu m$ and thickness $d=0.15 \mu m$. The superconductor is in the
Meissner state, i.e. $L=0$.
}
\label{fig3}
\end{figure}

\begin{figure}
\caption{Phase diagram for the superconducting state of a ring of outer radius
$R_o/\xi = 4$ as function of the inner radius $R_i$. The transition between
states with different vorticity $L$ is given by the solid curves. 
The region where the multi-vortex state has the lowest energy is shown by the
shaded area. The thick solid curve gives the normal/superconducting phase boundary.
}
\label{fig4}
\end{figure}

\begin{figure}
\caption{The Cooper pair density for a ring with $R_o/\xi=4$ and $R_i/\xi=0.4$
for $H/H_{c2} = 0.72 (1), 0.795 (2), 0.87 (3), 0.945 (4)$.
}
\label{fig5} 
\end{figure}

\begin{figure}
\caption{The magnetic field distribution and a contour plot of the phase of the
superconducting wavefunction for a ring with $R_o/\xi=4$, $R_i/\xi=1$ and
$d/\xi = 0.005$ in 
the presence of an external magnetic field of $H/H_{c2}= 1.145$.
}
\label{fig6}
\end{figure}

\begin{figure}
\caption{
(a) The flux through the hole of a ring with $R_o=2 \mu m$ and $R_i=1 \mu m$ as
function of the magnetic field. (b) The effective radius for the circular surface area
through which flux is quantized. 
}
\label{fig7}
\end{figure}

\begin{figure}
\caption{
(a) The radial dependence of the superconducting current through the ring of
Fig. 7 and (b) the corresponding flux through the area with radius $\rho$.
}
\label{fig8}
\end{figure}

\begin{figure}
\caption{
The magnetic field range ($\Delta \phi = \pi R^2_o \Delta H$) over which the
L-state stays in the ground state for different values of the inner radius $R_i$
and for outer radius (a) $R_o/\xi=2$ and (b) $R_o/\xi=4$. 
The horizontal dashed lines in (a) are for an increase with an integer number
of flux quanta through an area with radius $\sqrt{R_iR_o}$.
}
\label{fig9}
\end{figure}

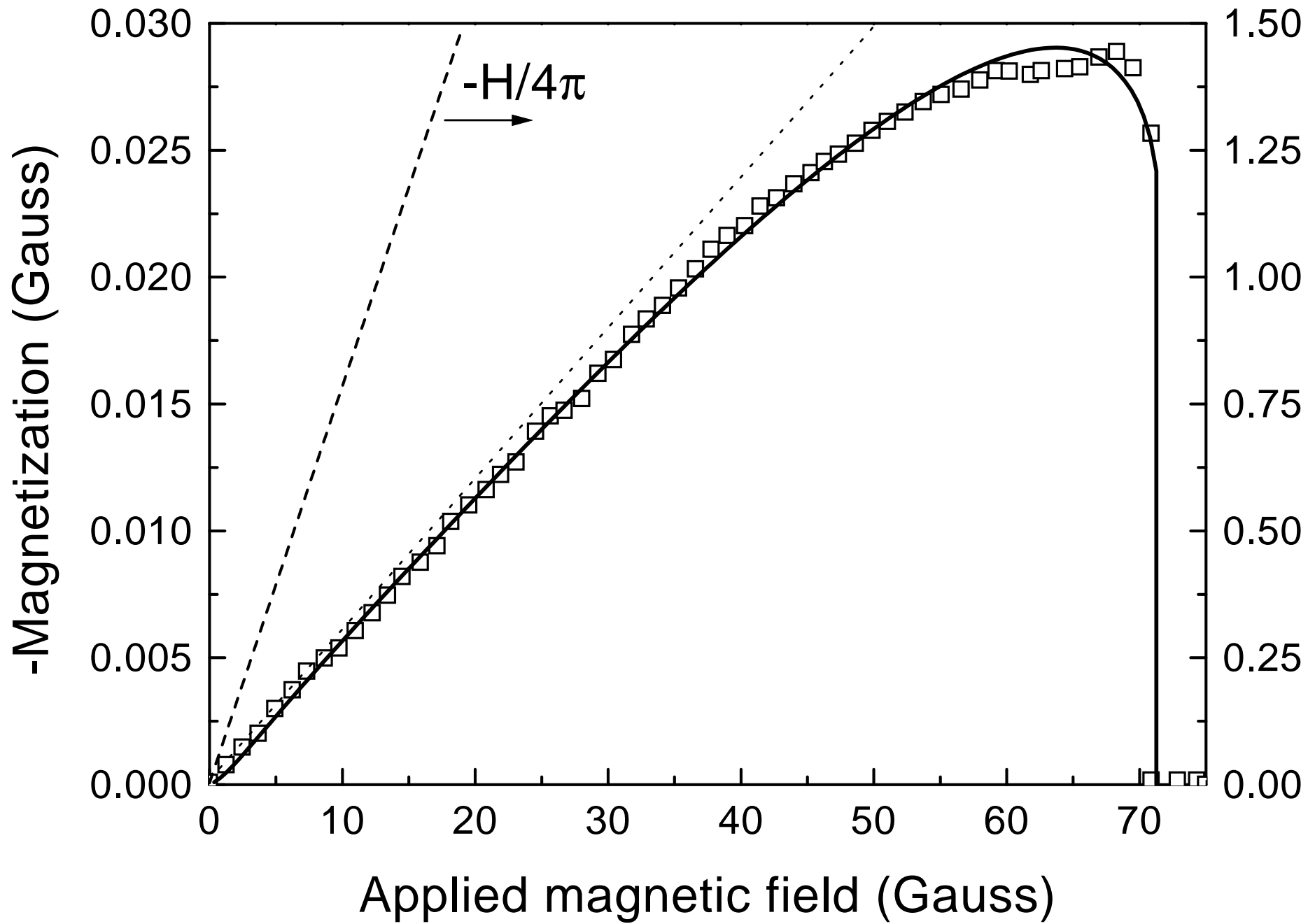
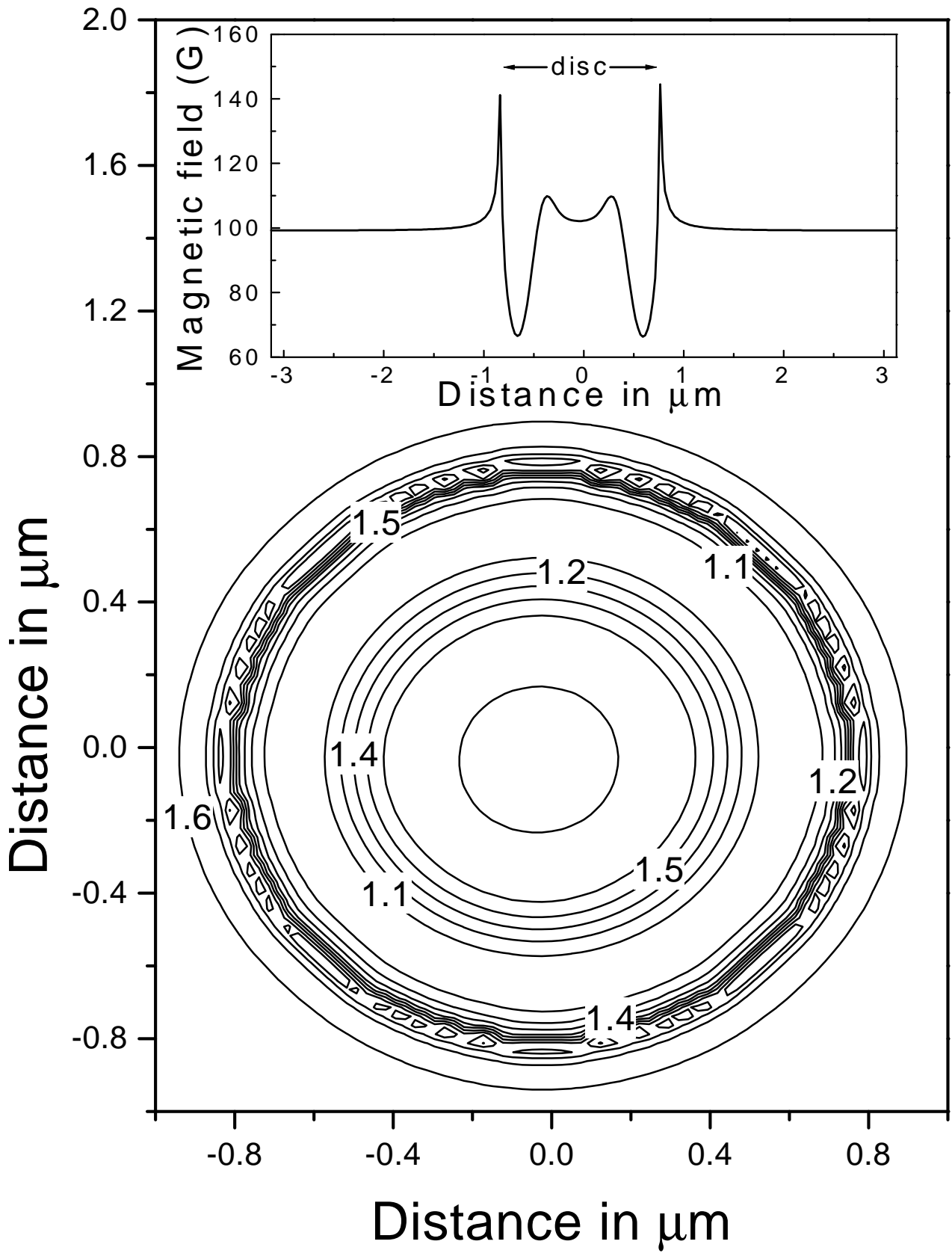
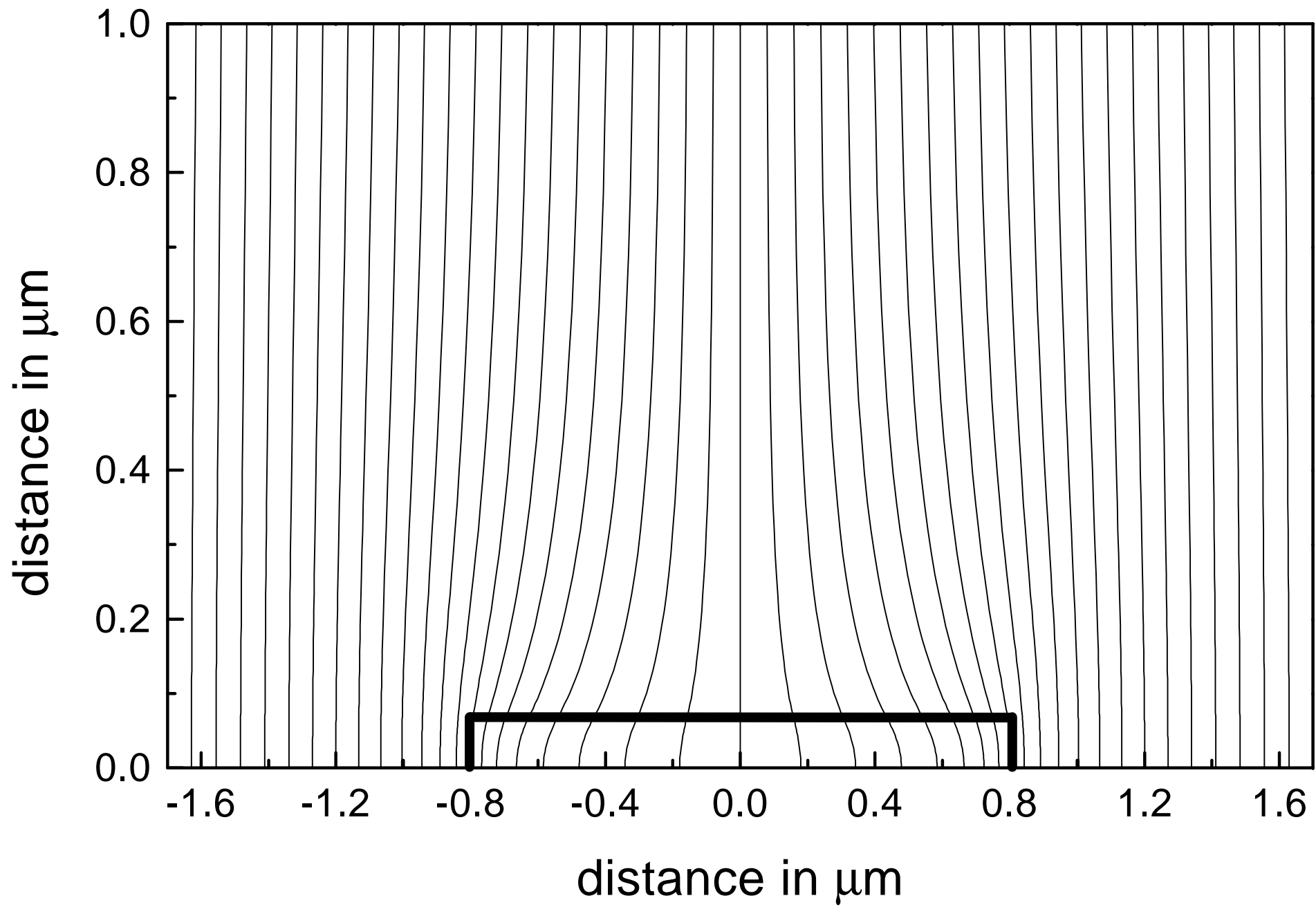
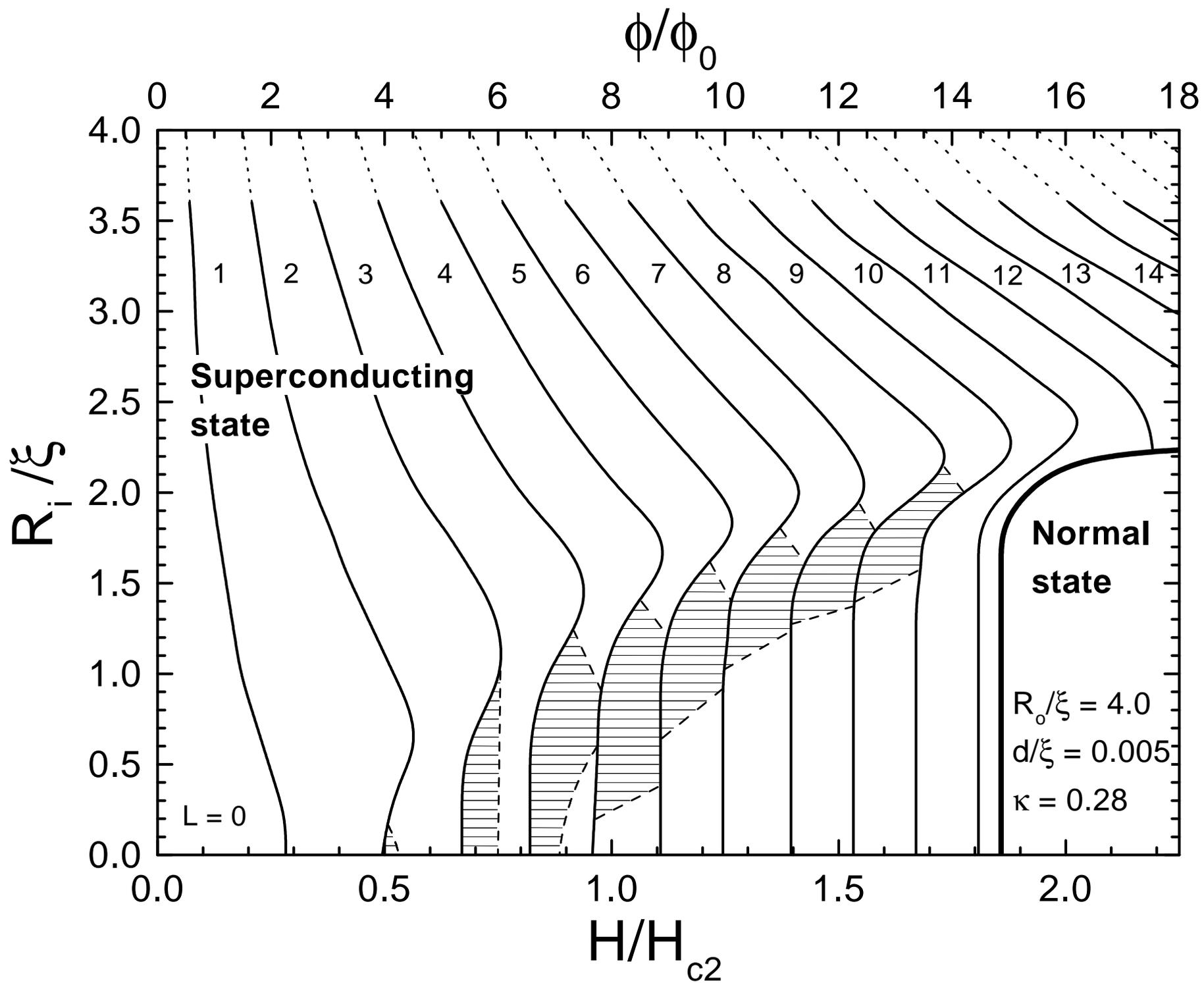
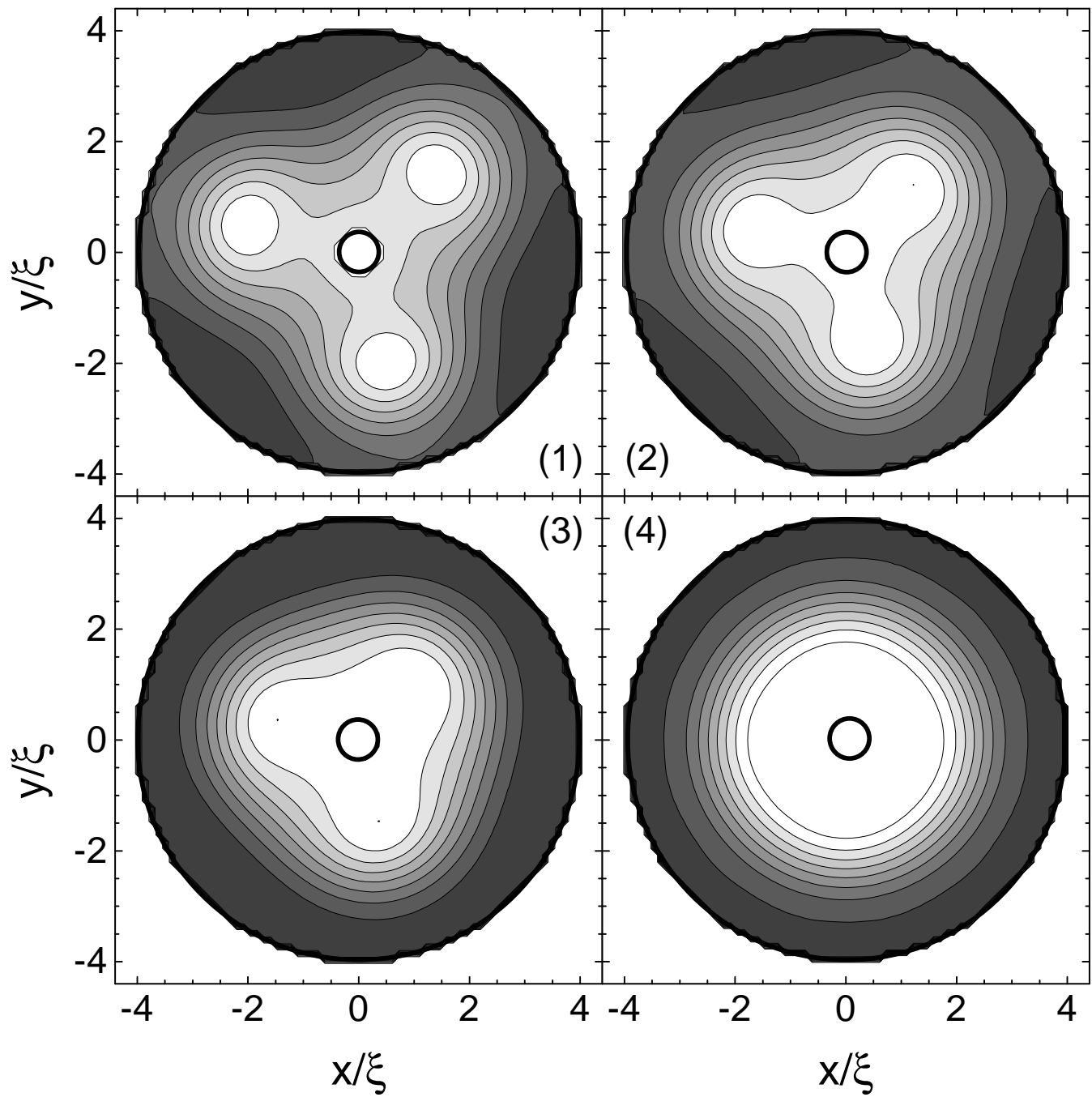
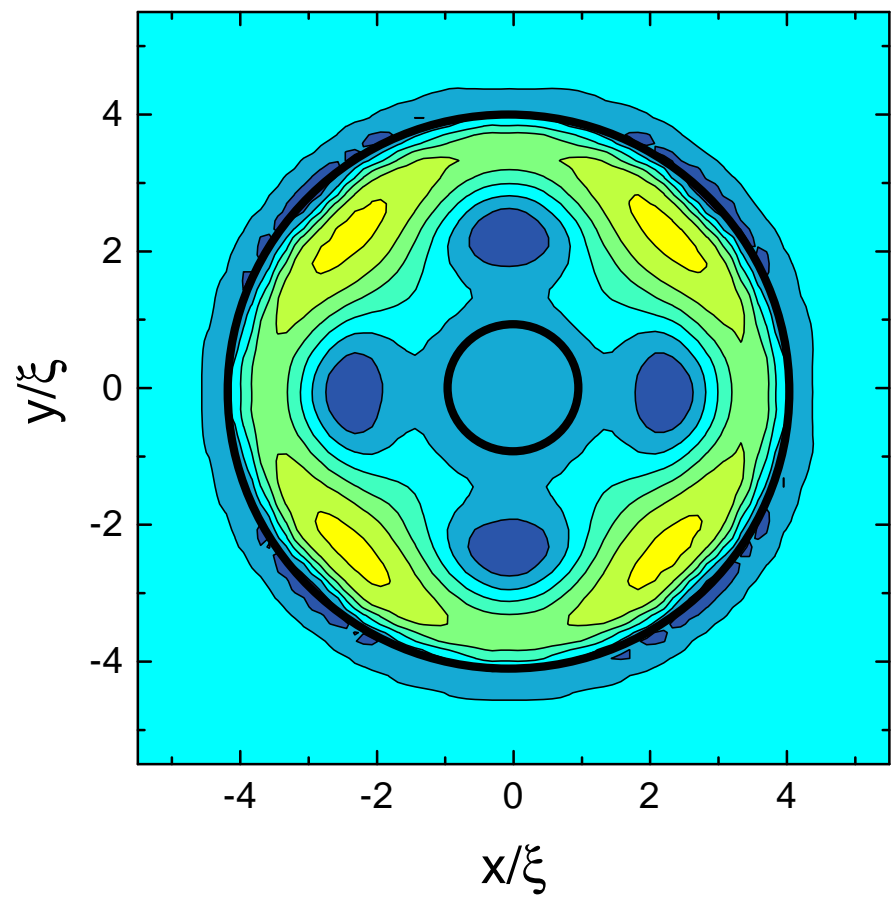
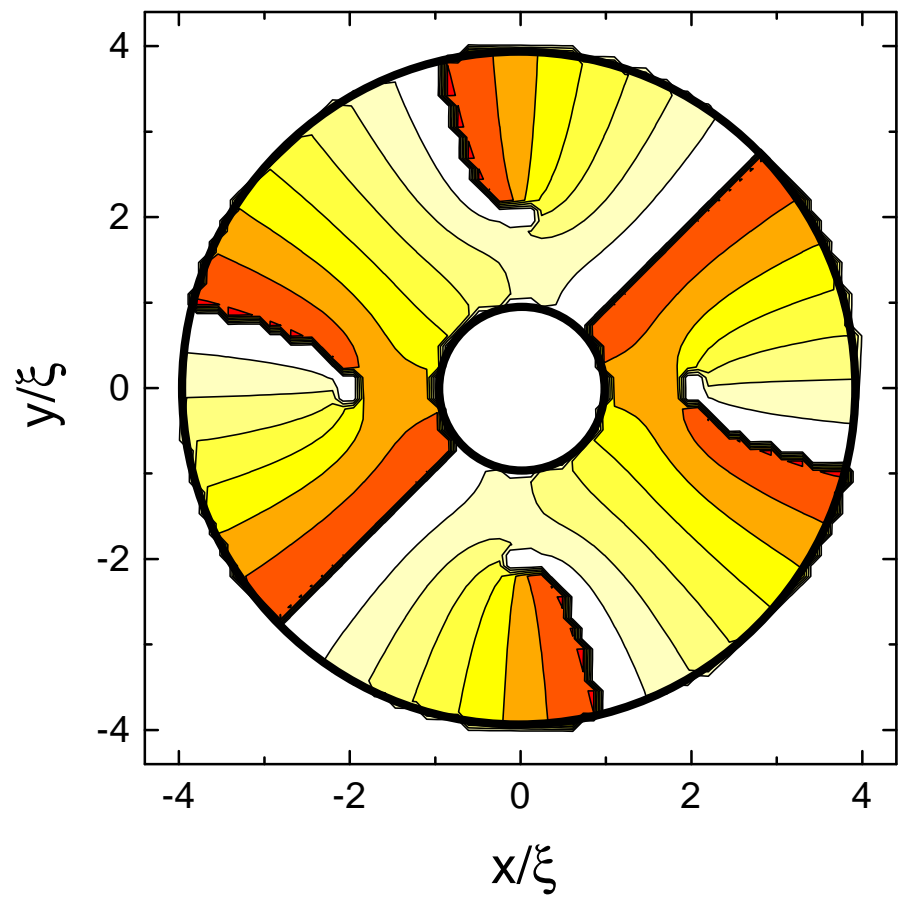
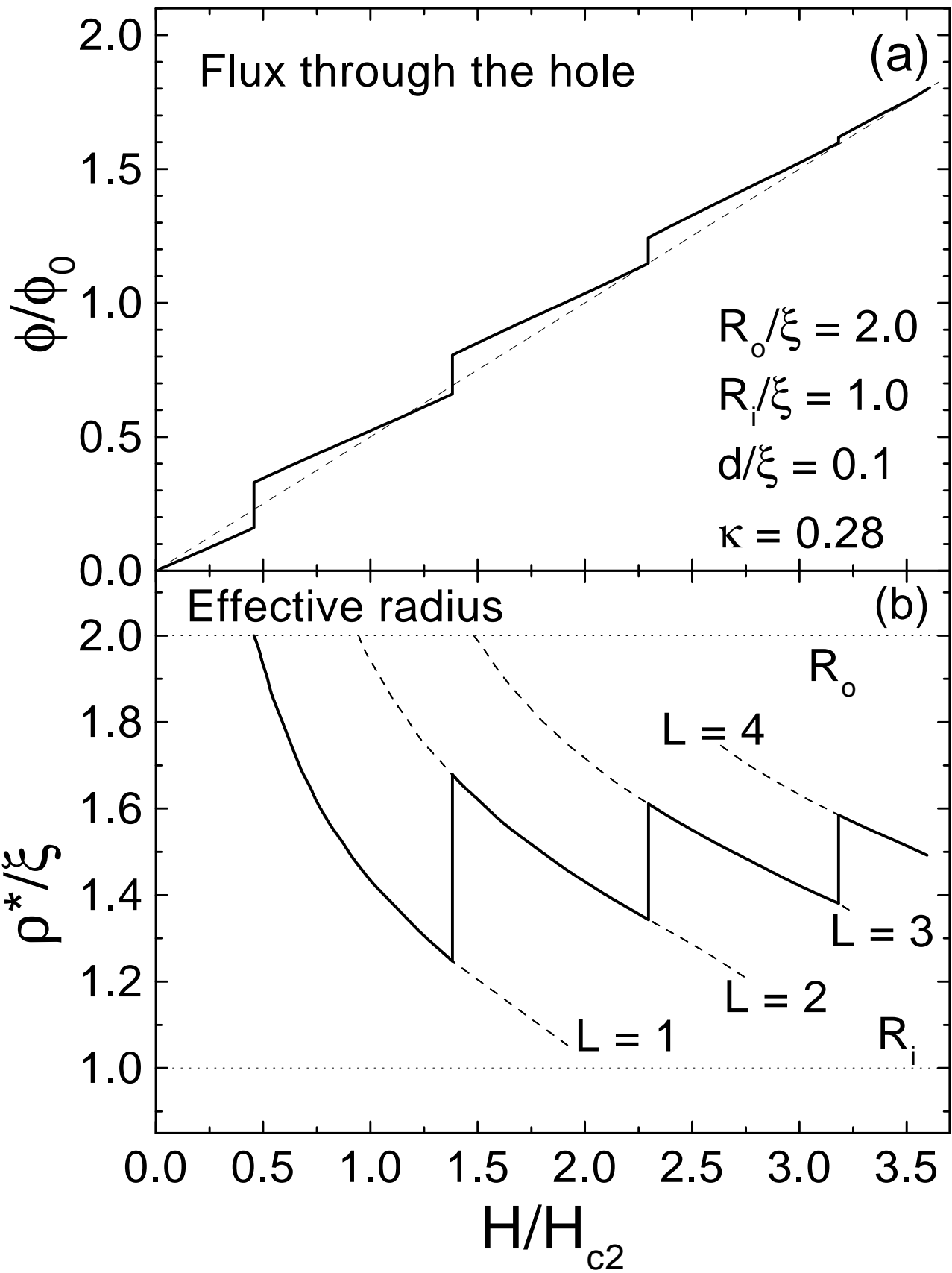
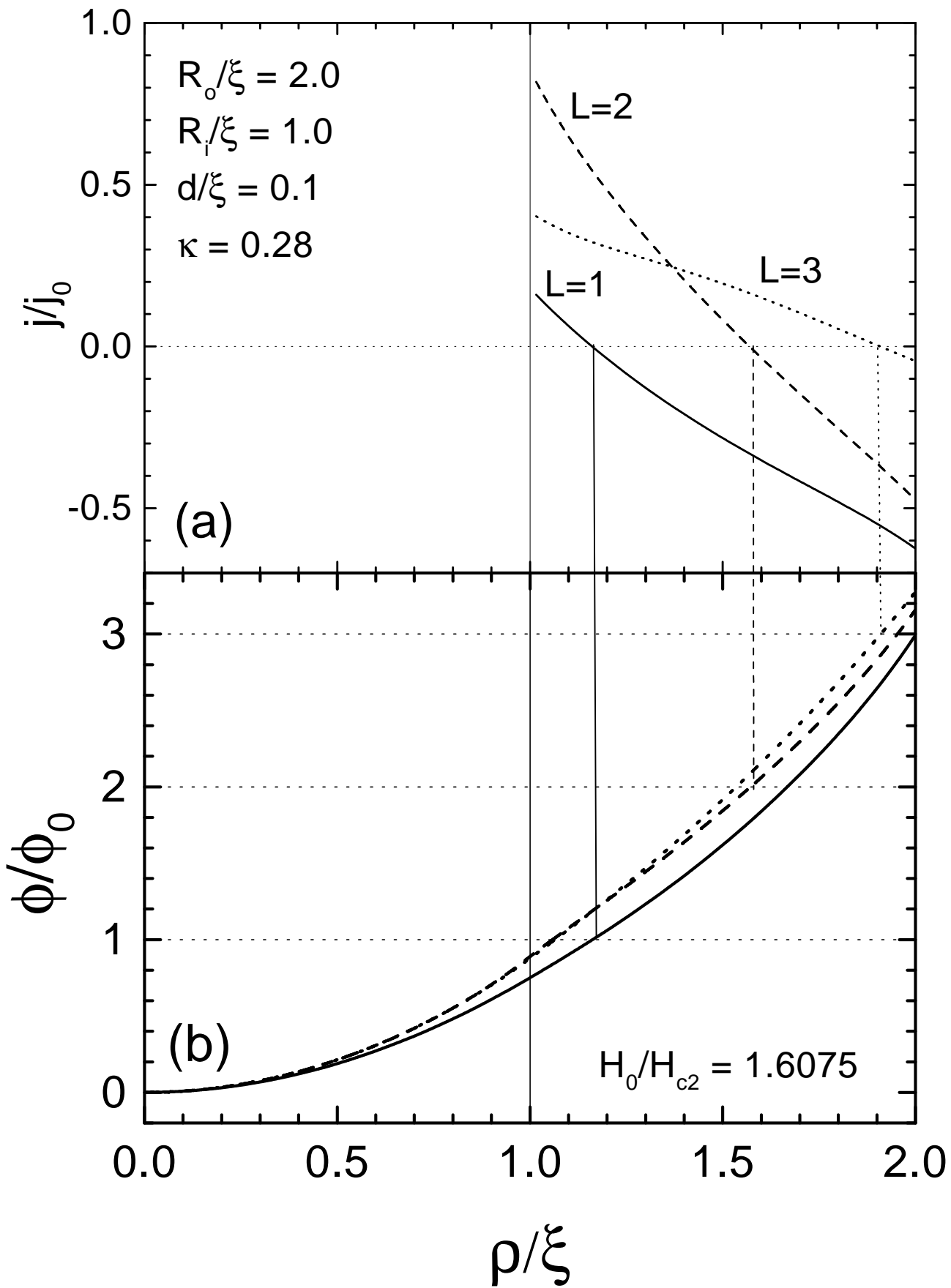
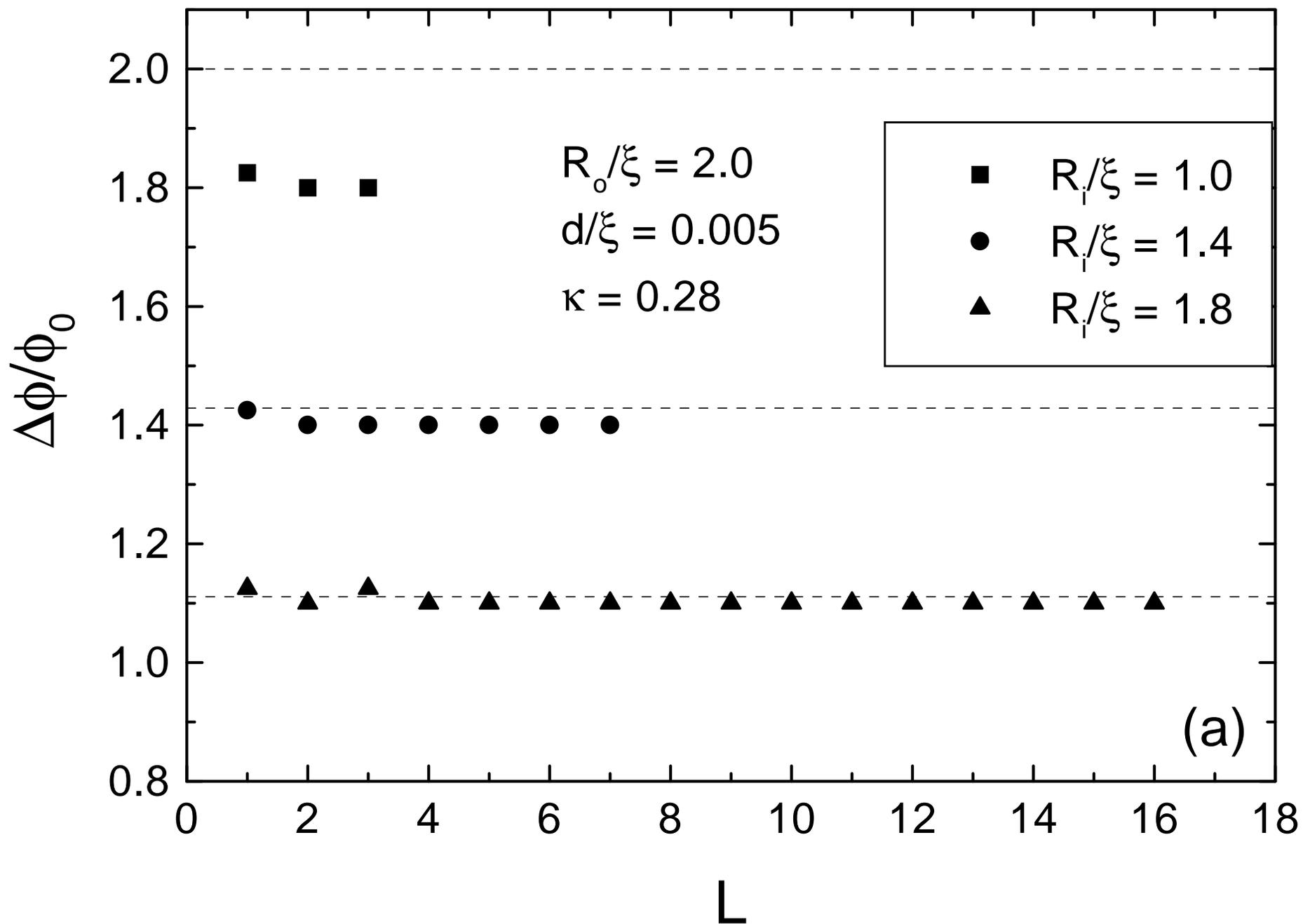
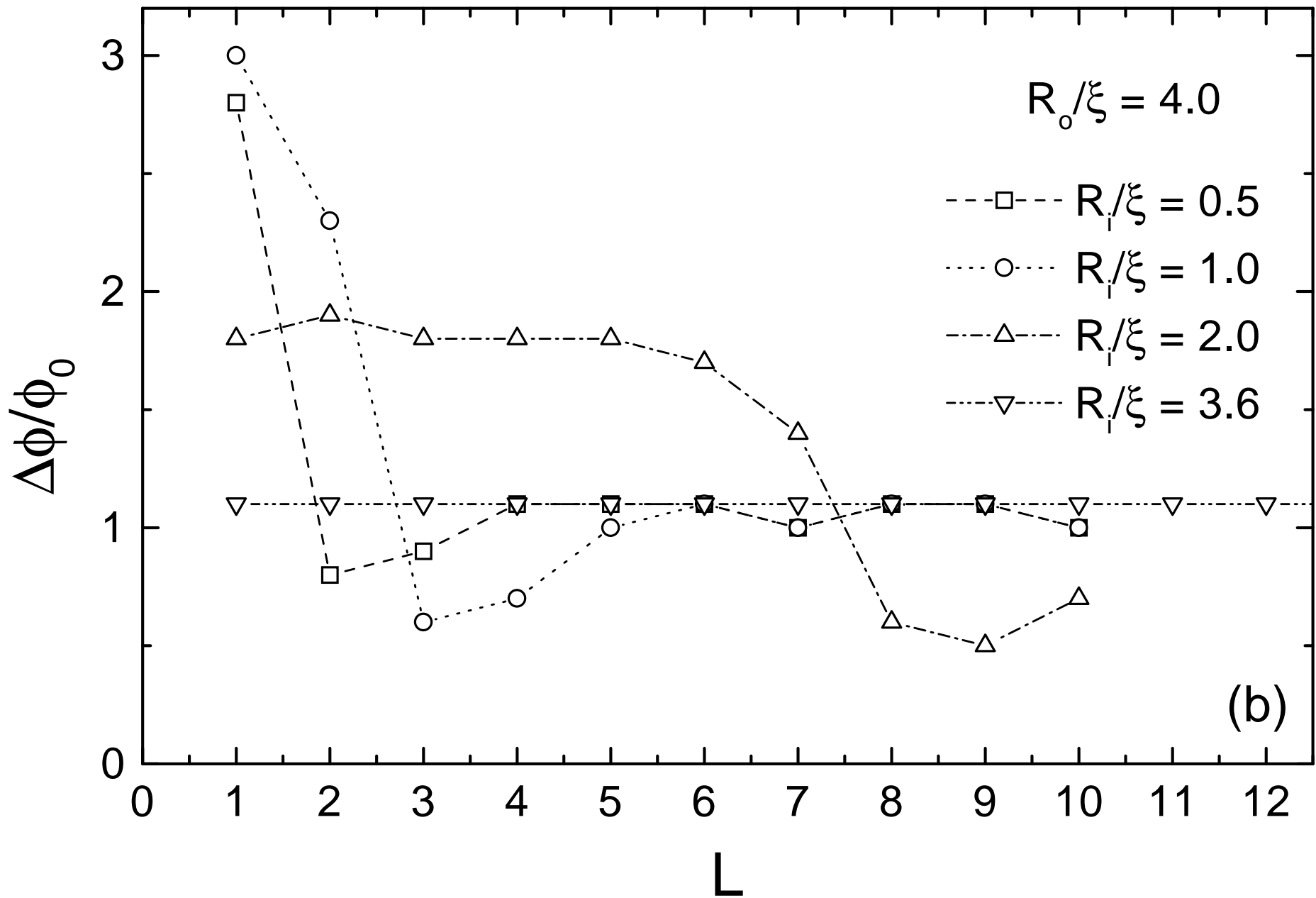
\end{document}